\documentclass[showpacs,amsmath,amssymb,superscriptaddress,twocolumn,prl]{revtex4}
\usepackage{graphicx}% Include figure files

\begin{document}
%Basic ideas:
% - They claim: systematic improvement; non-global comms; S inversion
% - But: Conquest, ONETEP, FEMTECK at least give system (with CP2K
%   offering another possibility)
% - Non-global comms is standard part of parallel, O(N): Challacombe,
%   Conquest and ONETEP (at least - maybe also Rudberg)
% - S inversion comes from Stechel in 1994; Hotelling also gives good
%   route; OMM methods are important in this area
% - Scalability - refer to Conquest million atoms, and K computer
% - Also note that they don't actually show MD results, and that this
%   is challenging (cf TC methods and XBOMD Lagrangian from Niklasson)
\title{Comment on ``Accurate and Scalable O(N) Algorithm for
  First-Principles Molecular-Dynamics Computations on Large Parallel Computers''}

\author{David \surname{Bowler}}
    \email{david.bowler@ucl.ac.uk}
	\affiliation{London Centre for Nanotechnology, 17-19 Gordon St, London, WC1H 0AH, U.K.} 
	\affiliation{Thomas Young Centre and Department of Physics
          $\&$ Astronomy, UCL, Gower St, London, WC1E 6BT, U.K.} 
	\affiliation{MANA, National Institute for Materials
          Science, 1-1 Namiki, Tsukuba, Ibaraki, 305-0044 Japan}
\author{Tsuyoshi Miyazaki}
%\email{MIYAZAKI.Tsuyoshi@nims.go.jp}
\affiliation{National Institute for Materials Science, 1-2-1 Sengen,
  Tsukuba, Ibaraki, 305-0047 Japan}
\author{Lionel A. Truflandier}
%\email{l.truflandier@ism.u-bordeaux1.fr}
\affiliation{ISM, Universit{\'e} Bordeaux I, 351 Cours de la Lib{\'e}ration, 33405 Talence, France}
\author{Michael J. Gillan}
%\email{m.gillan@ucl.ac.uk}
	\affiliation{London Centre for Nanotechnology, 17-19 Gordon St, London, WC1H 0AH, U.K.} 
	\affiliation{Thomas Young Centre and Department of Physics
          $\&$ Astronomy, UCL, Gower St, London, WC1E 6BT, U.K.} 

\maketitle

While we acknowledge the progress made by Osei-Kuffuor and Fattebert
in developing their O(N) algorithm\cite{Osei-Kuffuor:2014si}, 
we disagree with a number of their claims and
statement, in
particular that they have presented the first truly scalable O(N)
molecular dynamics algorithm.  The claims we will discuss are:
controllable accuracy; non-global communications and 
scalability; inversion of the overlap matrix; and the
lack of actual molecular dynamics in their results.

There are a number of O(N) codes already available which offer
controllable accuracy in the basis set.  ONETEP\cite{Skylaris:2005ai} uses periodic sinc 
functions\cite{Mostofi:2002wb}, while \textsc{Conquest}\cite{Bowler:2002pt} uses b-spline
functions\cite{Hernandez:1997ay} and FEMTECK\cite{Tsuchida:2007kq} uses finite elements.\cite{Tsuchida:1998dp}.
In all these codes, the accuracy is systematically controlled using a
grid spacing which is directly equivalent to a plane-wave cutoff, and
involves no approximation in the kinetic energy (whereas a finite
difference approach approximates the kinetic
energy\cite{Skylaris:2001gs}).  We also note that, while pseudo-atomic
orbitals and gaussian orbitals are hard to converge systematically, it
is possible to do so\cite{Dunning:2001rt,Blum:2009kw}.

Despite the authors' assertion that most O(N) algorithms do not remove
global communications, the principle of removing global communications to achieve scalability
is well established in the O(N) field.  This is shown by papers on sparse matrix
multiplication\cite{Saravanan:2003ye,Rubensson:2011mi} and sparse, parallel matrix multiplication in FreeON\cite{Challacombe:2000vh,Challacombe:2010pd},
ONETEP\cite{Hine:2010fj}, CP2K\cite{VandeVondele:2012qe} and
\textsc{Conquest}\cite{Bowler:2001yo}. 

We disagree with the authors'
assertion that "the parallel implementation of [algorithms to invert
the S matrix] generally require some global coupling".  There are a
number of existing approaches to inverting the S matrix, including
the orbital minimisation method
(OMM)\cite{Mauri:1993lf}\cite{Ordejon:1993zk} used by Siesta and FEMTECK, the
method used in OpenMX\cite{Ozaki:2001dd}, and
Hotelling's or Schultz's method used by \textsc{Conquest} and
ONETEP (this method is scalable and O(N) with
sparse matrix algebra), as well as the approximate inverse methods cited
by the authors.  We note that the approach that the authors suggest
has already been proposed\cite{Stechel:1994mk}.

\begin{figure}[h]
  \centering
  \includegraphics[width=\columnwidth]{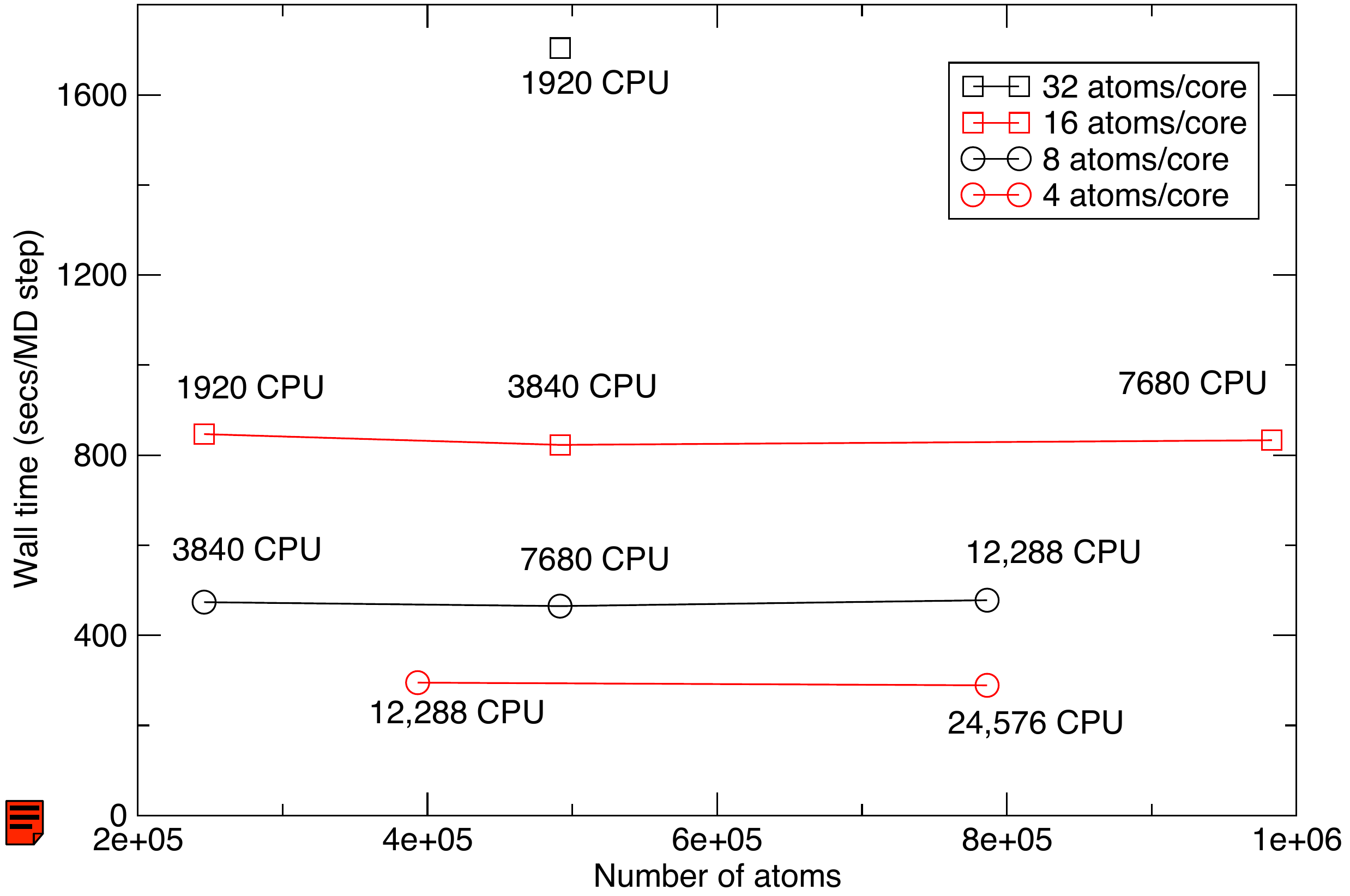}%WeakScalingNew_Ver5}
  \caption{Weak scaling for CONQUEST on the K computer, showing
    scaling to around 200,000 cores (8 cores per CPU).}
  \label{fig:weak}
\end{figure}

Moreover, it is important to recall that one of the major efforts in
the O(N) community in recent years has been to develop locally
communicating, scalable codes.  CP2K has 
demonstrated calculations on 1,000,000 atoms with density functional
tight binding (DFTB) and 96,000 water molecules with DFT, scaling to
46656 cores\cite{VandeVondele:2012qe}.
%6000 water with DFT and 1000 cores\cite{Khaliullin:2013cs}
\textsc{Conquest} has demonstrated scaling to over 2,000,000 atoms\cite{Bowler:2010uq}
on 4,096 processors, and recently scaled to 196,000 cores on the K
computer\cite{Miyazaki:2013ng} as shown in Fig.~\ref{fig:weak}; while
the data presented in this figure used pseudo-atomic orbitals as the
basis, the same scaling is found for the systematically-improvable
blip basis set (the blip basis set will be approximately
10 times faster than the DZP basis set used in this calculation,
giving times of half a minute per MD step or below).  In the data in Fig. 2 presented
by Osei-Kuffuor and Fattebert, there is a slow increase
in wall clock time with system size on the IBM BGQ which indicates some residual
problems with scalability in their implementation.

The accurate, efficient calculation of molecular dynamics with O(N)
complexity is a significant challenge, with problems including energy
conservation and efficiency of re-use of electronic
structure\cite{Niklasson:2006uk} and efficient load balancing.  We note that,
despite the title of the paper,  the authors
do not actually show any results from molecular dynamics, while these
have been demonstrated elsewhere, e.g. MD on 1,000 atoms of
ethanol (though using direct inversion of S, which is not
scalable)\cite{Tsuchida:2008ai}, and accurate, energy conserving O(N)
MD\cite{Cawkwell:2012rw}(without testing scalability). We have
demonstrated efficient relaxation of large nanostructures\cite{Miyazaki:2008kx} with
systems of 23,000 atoms, recently extended to over 100,000 atoms, and
have recently implemented efficient, energy conserving molecular
dynamics in Conquest\cite{Arita:2014rq}.

%In all our comments we simply want to note that there have been other
%demonstrations of scalable, efficient linear scaling approaches to
%electronic structure calculations including 
%\emph{ab initio} molecular dynamics, which should have been referenced
%and acknowledged by the authors of the present paper.

%\bibliography{tag_O(N),other}

\begin{thebibliography}{29}
\expandafter\ifx\csname natexlab\endcsname\relax\def\natexlab#1{#1}\fi
\expandafter\ifx\csname bibnamefont\endcsname\relax
  \def\bibnamefont#1{#1}\fi
\expandafter\ifx\csname bibfnamefont\endcsname\relax
  \def\bibfnamefont#1{#1}\fi
\expandafter\ifx\csname citenamefont\endcsname\relax
  \def\citenamefont#1{#1}\fi
\expandafter\ifx\csname url\endcsname\relax
  \def\url#1{\texttt{#1}}\fi
\expandafter\ifx\csname urlprefix\endcsname\relax\def\urlprefix{URL }\fi
\providecommand{\bibinfo}[2]{#2}
\providecommand{\eprint}[2][]{\url{#2}}

\bibitem[{\citenamefont{Osei-Kuffuor and
  Fattebert}(2014)}]{Osei-Kuffuor:2014si}
\bibinfo{author}{\bibfnamefont{D.}~\bibnamefont{Osei-Kuffuor}}
  \bibnamefont{and} \bibinfo{author}{\bibfnamefont{J.-L.}
  \bibnamefont{Fattebert}}, \bibinfo{journal}{Phys. Rev. Lett.}
  \textbf{\bibinfo{volume}{112}}, \bibinfo{pages}{046401}
  (\bibinfo{year}{2014}).

\bibitem[{\citenamefont{Skylaris et~al.}(2005)\citenamefont{Skylaris, Haynes,
  Mostofi, and Payne}}]{Skylaris:2005ai}
\bibinfo{author}{\bibfnamefont{C.-K.} \bibnamefont{Skylaris}},
  \bibinfo{author}{\bibfnamefont{P.~D.} \bibnamefont{Haynes}},
  \bibinfo{author}{\bibfnamefont{A.~A.} \bibnamefont{Mostofi}},
  \bibnamefont{and} \bibinfo{author}{\bibfnamefont{M.~C.} \bibnamefont{Payne}},
  \bibinfo{journal}{J. Chem. Phys.} \textbf{\bibinfo{volume}{122}},
  \bibinfo{eid}{084119} (\bibinfo{year}{2005}).

\bibitem[{\citenamefont{Mostofi et~al.}(2002)\citenamefont{Mostofi, Skylaris,
  Haynes, and Payne}}]{Mostofi:2002wb}
\bibinfo{author}{\bibfnamefont{A.~A.} \bibnamefont{Mostofi}},
  \bibinfo{author}{\bibfnamefont{C.-K.} \bibnamefont{Skylaris}},
  \bibinfo{author}{\bibfnamefont{P.~D.} \bibnamefont{Haynes}},
  \bibnamefont{and} \bibinfo{author}{\bibfnamefont{M.~C.} \bibnamefont{Payne}},
  \bibinfo{journal}{Comp. Phys. Commun.} \textbf{\bibinfo{volume}{147}},
  \bibinfo{pages}{788} (\bibinfo{year}{2002}).

\bibitem[{\citenamefont{Bowler et~al.}(2002)\citenamefont{Bowler, Miyazaki, and
  Gillan}}]{Bowler:2002pt}
\bibinfo{author}{\bibfnamefont{D.~R.} \bibnamefont{Bowler}},
  \bibinfo{author}{\bibfnamefont{T.}~\bibnamefont{Miyazaki}}, \bibnamefont{and}
  \bibinfo{author}{\bibfnamefont{M.~J.} \bibnamefont{Gillan}},
  \bibinfo{journal}{J. Phys.: Condens. Matter} \textbf{\bibinfo{volume}{14}},
  \bibinfo{pages}{2781} (\bibinfo{year}{2002}).

%\bibitem[{\citenamefont{Bowler et~al.}(2006)\citenamefont{Bowler, Choudhury,
%  Gillan, and Miyazaki}}]{Bowler:2006wa}
%\bibinfo{author}{\bibfnamefont{D.~R.} \bibnamefont{Bowler}},
%  \bibinfo{author}{\bibfnamefont{R.}~\bibnamefont{Choudhury}},
%  \bibinfo{author}{\bibfnamefont{M.~J.} \bibnamefont{Gillan}},
%  \bibnamefont{and} \bibinfo{author}{\bibfnamefont{T.}~\bibnamefont{Miyazaki}},
%  \bibinfo{journal}{phys. stat. sol. b} \textbf{\bibinfo{volume}{243}},
%  \bibinfo{pages}{989} (\bibinfo{year}{2006}).
%
\bibitem[{\citenamefont{Hern\'andez et~al.}(1997)\citenamefont{Hern\'andez,
  Gillan, and Goringe}}]{Hernandez:1997ay}
\bibinfo{author}{\bibfnamefont{E.}~\bibnamefont{Hern\'andez}},
  \bibinfo{author}{\bibfnamefont{M.~J.} \bibnamefont{Gillan}},
  \bibnamefont{and} \bibinfo{author}{\bibfnamefont{C.~M.}
  \bibnamefont{Goringe}}, \bibinfo{journal}{Phys. Rev. B}
  \textbf{\bibinfo{volume}{55}}, \bibinfo{pages}{13485} (\bibinfo{year}{1997}).

\bibitem[{\citenamefont{Tsuchida}(2007)}]{Tsuchida:2007kq}
\bibinfo{author}{\bibfnamefont{E.}~\bibnamefont{Tsuchida}},
  \bibinfo{journal}{J. Phys. Soc. Japan} \textbf{\bibinfo{volume}{76}},
  \bibinfo{pages}{034708} (\bibinfo{year}{2007}).

\bibitem[{\citenamefont{Tsuchida and Tsukada}(1998)}]{Tsuchida:1998dp}
\bibinfo{author}{\bibfnamefont{E.}~\bibnamefont{Tsuchida}} \bibnamefont{and}
  \bibinfo{author}{\bibfnamefont{M.}~\bibnamefont{Tsukada}},
  \bibinfo{journal}{J. Phys. Soc. Japan} \textbf{\bibinfo{volume}{67}},
  \bibinfo{pages}{3844} (\bibinfo{year}{1998}).

\bibitem[{\citenamefont{Skylaris et~al.}(2001)\citenamefont{Skylaris, Mostofi,
  Haynes, Pickard, and Payne}}]{Skylaris:2001gs}
\bibinfo{author}{\bibfnamefont{C.-K.} \bibnamefont{Skylaris}},
  \bibinfo{author}{\bibfnamefont{A.~A.} \bibnamefont{Mostofi}},
  \bibinfo{author}{\bibfnamefont{P.~D.} \bibnamefont{Haynes}},
  \bibinfo{author}{\bibfnamefont{C.~J.} \bibnamefont{Pickard}},
  \bibnamefont{and} \bibinfo{author}{\bibfnamefont{M.~C.} \bibnamefont{Payne}},
  \bibinfo{journal}{Comp. Phys. Commun.} \textbf{\bibinfo{volume}{140}},
  \bibinfo{pages}{315 } (\bibinfo{year}{2001}).

\bibitem[{\citenamefont{Dunning et~al.}(2001)\citenamefont{Dunning, Peterson,
  and Wilson}}]{Dunning:2001rt}
\bibinfo{author}{\bibfnamefont{T.~H.} \bibnamefont{Dunning},
  \bibfnamefont{Jr.}}, \bibinfo{author}{\bibfnamefont{K.~A.}
  \bibnamefont{Peterson}}, \bibnamefont{and}
  \bibinfo{author}{\bibfnamefont{A.~K.} \bibnamefont{Wilson}},
  \bibinfo{journal}{J. Chem. Phys.} \textbf{\bibinfo{volume}{114}},
  \bibinfo{pages}{9244} (\bibinfo{year}{2001}).

\bibitem[{\citenamefont{Blum et~al.}(2009)\citenamefont{Blum, Gehrke, Hanke,
  Havu, Havu, Ren, Reuter, and Scheffler}}]{Blum:2009kw}
\bibinfo{author}{\bibfnamefont{V.}~\bibnamefont{Blum}},
  \bibinfo{author}{\bibfnamefont{R.}~\bibnamefont{Gehrke}},
  \bibinfo{author}{\bibfnamefont{F.}~\bibnamefont{Hanke}},
  \bibinfo{author}{\bibfnamefont{P.}~\bibnamefont{Havu}},
  \bibinfo{author}{\bibfnamefont{V.}~\bibnamefont{Havu}},
  \bibinfo{author}{\bibfnamefont{X.}~\bibnamefont{Ren}},
  \bibinfo{author}{\bibfnamefont{K.}~\bibnamefont{Reuter}}, \bibnamefont{and}
  \bibinfo{author}{\bibfnamefont{M.}~\bibnamefont{Scheffler}},
  \bibinfo{journal}{Comp. Phys. Commun.} \textbf{\bibinfo{volume}{180}},
  \bibinfo{pages}{2175} (\bibinfo{year}{2009}).

\bibitem[{\citenamefont{Saravanan et~al.}(2003)\citenamefont{Saravanan, Shao,
  Baer, Ross, and Head-Gordon}}]{Saravanan:2003ye}
\bibinfo{author}{\bibfnamefont{C.}~\bibnamefont{Saravanan}},
  \bibinfo{author}{\bibfnamefont{Y.}~\bibnamefont{Shao}},
  \bibinfo{author}{\bibfnamefont{R.}~\bibnamefont{Baer}},
  \bibinfo{author}{\bibfnamefont{P.~N.} \bibnamefont{Ross}}, \bibnamefont{and}
  \bibinfo{author}{\bibfnamefont{M.}~\bibnamefont{Head-Gordon}},
  \bibinfo{journal}{J. Comput. Chem.} \textbf{\bibinfo{volume}{24}},
  \bibinfo{pages}{618} (\bibinfo{year}{2003}).

\bibitem[{\citenamefont{Rubensson and Rudberg}(2011)}]{Rubensson:2011mi}
\bibinfo{author}{\bibfnamefont{E.~H.} \bibnamefont{Rubensson}}
  \bibnamefont{and} \bibinfo{author}{\bibfnamefont{E.}~\bibnamefont{Rudberg}},
  \bibinfo{journal}{J. Comput. Chem.} \textbf{\bibinfo{volume}{32}},
  \bibinfo{pages}{1411} (\bibinfo{year}{2011}).

\bibitem[{\citenamefont{Challacombe}(2000)}]{Challacombe:2000vh}
\bibinfo{author}{\bibfnamefont{M.}~\bibnamefont{Challacombe}},
  \bibinfo{journal}{Comp. Phys. Commun.} \textbf{\bibinfo{volume}{128}},
  \bibinfo{pages}{93} (\bibinfo{year}{2000}).

\bibitem[{\citenamefont{Challacombe and Bock}(2010)}]{Challacombe:2010pd}
\bibinfo{author}{\bibfnamefont{M.}~\bibnamefont{Challacombe}} \bibnamefont{and}
  \bibinfo{author}{\bibfnamefont{N.}~\bibnamefont{Bock}},
  \bibinfo{journal}{arxiv} \textbf{\bibinfo{volume}{1011.3534}}
  (\bibinfo{year}{2010}).

\bibitem[{\citenamefont{Hine et~al.}(2010)\citenamefont{Hine, Haynes, Mostofi,
  and Payne}}]{Hine:2010fj}
\bibinfo{author}{\bibfnamefont{N.~D.~M.} \bibnamefont{Hine}},
  \bibinfo{author}{\bibfnamefont{P.~D.} \bibnamefont{Haynes}},
  \bibinfo{author}{\bibfnamefont{A.~A.} \bibnamefont{Mostofi}},
  \bibnamefont{and} \bibinfo{author}{\bibfnamefont{M.~C.} \bibnamefont{Payne}},
  \bibinfo{journal}{J. Chem. Phys.} \textbf{\bibinfo{volume}{133}},
  \bibinfo{eid}{114111} (\bibinfo{year}{2010}).

\bibitem[{\citenamefont{VandeVondele et~al.}(2012)\citenamefont{VandeVondele,
  Bor\v{s}tnik, and Hutter}}]{VandeVondele:2012qe}
\bibinfo{author}{\bibfnamefont{J.}~\bibnamefont{VandeVondele}},
  \bibinfo{author}{\bibfnamefont{U.}~\bibnamefont{Bor\v{s}tnik}},
  \bibnamefont{and} \bibinfo{author}{\bibfnamefont{J.}~\bibnamefont{Hutter}},
  \bibinfo{journal}{J. Chem. Theory Comput.} \textbf{\bibinfo{volume}{8}},
  \bibinfo{pages}{3565} (\bibinfo{year}{2012}).

\bibitem[{\citenamefont{Bowler et~al.}(2001)\citenamefont{Bowler, Miyazaki, and
  Gillan}}]{Bowler:2001yo}
\bibinfo{author}{\bibfnamefont{D.~R.} \bibnamefont{Bowler}},
  \bibinfo{author}{\bibfnamefont{T.}~\bibnamefont{Miyazaki}}, \bibnamefont{and}
  \bibinfo{author}{\bibfnamefont{M.~J.} \bibnamefont{Gillan}},
  \bibinfo{journal}{Comp. Phys. Commun.} \textbf{\bibinfo{volume}{137}},
  \bibinfo{pages}{255} (\bibinfo{year}{2001}).

\bibitem[{\citenamefont{Mauri et~al.}(1993)\citenamefont{Mauri, Galli, and
  Car}}]{Mauri:1993lf}
\bibinfo{author}{\bibfnamefont{F.}~\bibnamefont{Mauri}},
  \bibinfo{author}{\bibfnamefont{G.}~\bibnamefont{Galli}}, \bibnamefont{and}
  \bibinfo{author}{\bibfnamefont{R.}~\bibnamefont{Car}},
  \bibinfo{journal}{Phys. Rev. B} \textbf{\bibinfo{volume}{47}},
  \bibinfo{pages}{9973} (\bibinfo{year}{1993}).

\bibitem[{\citenamefont{Ordej\'on et~al.}(1993)\citenamefont{Ordej\'on,
  Drabold, Grumbach, and Martin}}]{Ordejon:1993zk}
\bibinfo{author}{\bibfnamefont{P.}~\bibnamefont{Ordej\'on}},
  \bibinfo{author}{\bibfnamefont{D.~A.} \bibnamefont{Drabold}},
  \bibinfo{author}{\bibfnamefont{M.~P.} \bibnamefont{Grumbach}},
  \bibnamefont{and} \bibinfo{author}{\bibfnamefont{R.~M.}
  \bibnamefont{Martin}}, \bibinfo{journal}{Phys. Rev. B}
  \textbf{\bibinfo{volume}{48}}, \bibinfo{pages}{14646} (\bibinfo{year}{1993}).

\bibitem[{\citenamefont{Ozaki}(2001)}]{Ozaki:2001dd}
\bibinfo{author}{\bibfnamefont{T.}~\bibnamefont{Ozaki}},
  \bibinfo{journal}{Phys. Rev. B} \textbf{\bibinfo{volume}{64}},
  \bibinfo{pages}{195110} (\bibinfo{year}{2001}).

\bibitem[{\citenamefont{Stechel et~al.}(1994)\citenamefont{Stechel, Williams,
  and Feibelman}}]{Stechel:1994mk}
\bibinfo{author}{\bibfnamefont{E.~B.} \bibnamefont{Stechel}},
  \bibinfo{author}{\bibfnamefont{A.~R.} \bibnamefont{Williams}},
  \bibnamefont{and} \bibinfo{author}{\bibfnamefont{P.~J.}
  \bibnamefont{Feibelman}}, \bibinfo{journal}{Phys. Rev. B}
  \textbf{\bibinfo{volume}{49}}, \bibinfo{pages}{10088} (\bibinfo{year}{1994}).

\bibitem[{\citenamefont{Bowler and Miyazaki}(2010)}]{Bowler:2010uq}
\bibinfo{author}{\bibfnamefont{D.~R.} \bibnamefont{Bowler}} \bibnamefont{and}
  \bibinfo{author}{\bibfnamefont{T.}~\bibnamefont{Miyazaki}},
  \bibinfo{journal}{J. Phys.: Condens. Matter} \textbf{\bibinfo{volume}{22}},
  \bibinfo{pages}{074207} (\bibinfo{year}{2010}).

\bibitem[{\citenamefont{Miyazaki}(2013)}]{Miyazaki:2013ng}
\bibinfo{author}{\bibfnamefont{T.}~\bibnamefont{Miyazaki}},
  \bibinfo{journal}{NIMS NOW} \textbf{\bibinfo{volume}{11}},
  \bibinfo{pages}{06} (\bibinfo{year}{2013}),
  \urlprefix\url{http://www.nims.go.jp/eng/publicity/nimsnow/2013/vol11\_09.html}.

\bibitem[{\citenamefont{Niklasson et~al.}(2006)\citenamefont{Niklasson,
  Tymczak, and Challacombe}}]{Niklasson:2006uk}
\bibinfo{author}{\bibfnamefont{A.~M.~N.} \bibnamefont{Niklasson}},
  \bibinfo{author}{\bibfnamefont{C.~J.} \bibnamefont{Tymczak}},
  \bibnamefont{and}
  \bibinfo{author}{\bibfnamefont{M.}~\bibnamefont{Challacombe}},
  \bibinfo{journal}{Phys. Rev. Lett.} \textbf{\bibinfo{volume}{97}},
  \bibinfo{pages}{123001} (\bibinfo{year}{2006}).

\bibitem[{\citenamefont{Tsuchida}(2008)}]{Tsuchida:2008ai}
\bibinfo{author}{\bibfnamefont{E.}~\bibnamefont{Tsuchida}},
  \bibinfo{journal}{J. Phys.: Condens. Matter} \textbf{\bibinfo{volume}{20}},
  \bibinfo{pages}{294212} (\bibinfo{year}{2008}).

\bibitem[{\citenamefont{Cawkwell and Niklasson}(2012)}]{Cawkwell:2012rw}
\bibinfo{author}{\bibfnamefont{M.~J.} \bibnamefont{Cawkwell}} \bibnamefont{and}
  \bibinfo{author}{\bibfnamefont{A.~M.~N.} \bibnamefont{Niklasson}},
  \bibinfo{journal}{J. Chem. Phys.} \textbf{\bibinfo{volume}{137}},
  \bibinfo{eid}{134105} (\bibinfo{year}{2012}).

\bibitem[{\citenamefont{Miyazaki et~al.}(2008)\citenamefont{Miyazaki, Bowler,
  Gillan, and Ohno}}]{Miyazaki:2008kx}
\bibinfo{author}{\bibfnamefont{T.}~\bibnamefont{Miyazaki}},
  \bibinfo{author}{\bibfnamefont{D.~R.} \bibnamefont{Bowler}},
  \bibinfo{author}{\bibfnamefont{M.~J.} \bibnamefont{Gillan}},
  \bibnamefont{and} \bibinfo{author}{\bibfnamefont{T.}~\bibnamefont{Ohno}},
  \bibinfo{journal}{J. Phys. Soc. Jpn.} \textbf{\bibinfo{volume}{77}},
  \bibinfo{pages}{123706} (\bibinfo{year}{2008}).

\bibitem[{\citenamefont{Arita et~al.}(2014)\citenamefont{Arita, Bowler, and
  Miyazaki}}]{Arita:2014rq}
\bibinfo{author}{\bibfnamefont{M.}~\bibnamefont{Arita}},
  \bibinfo{author}{\bibfnamefont{D.~R.} \bibnamefont{Bowler}},
  \bibnamefont{and} \bibinfo{author}{\bibfnamefont{T.}~\bibnamefont{Miyazaki}},
  \bibinfo{journal}{in preparation}  (\bibinfo{year}{2014}).

\end{thebibliography}

\end{document}